\documentclass[a4paper,fleqn,usenatbib]{mnras}

\usepackage[T1]{fontenc}
\usepackage{ae,aecompl}

\usepackage{graphicx}	
\usepackage{amsmath}	
\usepackage{amssymb}	
\usepackage{subfig}

\usepackage[normalem]{ulem}

\title[Core formation for frequent DM self-interactions]{Simulations of core formation for frequent dark matter self-interactions \thanks{Preprint numbers: DESY-19-020, TTK-19-06}}

\author[J.~Kummer, M.~Br\"uggen, K.~Dolag, F.~Kahlhoefer and K.~Schmidt-Hoberg
]
  {Janis Kummer$^{1,2}$\textsuperscript{\thanks{janis.kummer@desy.de}},
  Marcus Br\"uggen$^2$, Klaus Dolag$^3$, Felix Kahlhoefer$^4$ \newauthor and Kai Schmidt-Hoberg$^1$
\\
$^1$ DESY, Notkestrasse 85, D-22607 Hamburg, Germany\\
$^2$ Hamburger Sternwarte, Gojenbergsweg 112, D-21029 Hamburg, Germany\\
$^3$ Universit\"atssternwarte M\"unchen, Scheinerstrasse 1, D-81679 M\"unchen, Germany\\
$^4$ Institute for Theoretical Particle Physics and Cosmology (TTK), RWTH Aachen University, D-52056 Aachen, Germany\\
}

\date{Accepted XXX. Received YYY; in original form ZZZ}

\pubyear{2019}
\begin{document}
\label{firstpage}
\pagerange{\pageref{firstpage}--\pageref{lastpage}}
\maketitle

\begin{abstract}
We present the first $N$-body simulations that adapt the equations of smoothed particle hydrodynamics to capture the effect of dark matter self-interactions which are too frequent to be resolved explicitly. The relevant energy transfer equations are derived, the appropriate thermal conductivity is determined and the effects of different smoothing kernels are studied.
We apply our framework to simulate the formation of isothermal cores in isolated dark matter haloes and determine the core growth rate as a function of the self-scattering cross section. Our approach may be combined with explicit simulations of rare scatterings in order to simulate accurately the effects of arbitrary dark matter self-interactions in future cosmological simulations.
\end{abstract}



\section{Introduction}

The $\Lambda$CDM model based on the standard paradigm of collisionless cold dark matter (CDM) predicts the large-scale structure of the Universe remarkably well. But on small scales predictions from simulations of cosmological structure formation and observations appear to disagree with each other. In particular, pure CDM simulations predict cuspy density profiles with a steep slope ($\rho \propto r^{-1}$) in the centre of dark matter (DM) haloes~\citep{Dubinski:1991bm,Navarro:1995iw,Navarro:1996gj}, while observations tend to prefer more shallow profiles ($\rho \propto r^{0}$), corresponding to a central core~\citep{Flores:1994gz,Moore:1994yx,Moore:1999gc}. This disagreement is known as the {\it core-cusp} problem. 
The observed number of subhaloes in the Milky Way is by far smaller than predicted by simulations~\citep{Moore:1999nt,Klypin:1999uc}. This discrepancy is known as the {\it missing satellites} problem. Another disagreement called the {\it too-big-to-fail} problem states that the most massive subhaloes in CDM simulations are too dense in the centre to host the observed satellites of the Milky Way~\citep{BoylanKolchin:2011de,BoylanKolchin:2011dk}. Moreover, CDM is not able to explain the diversity of observed rotation curves of DM haloes with the same maximum circular velocity~\citep{Oman:2015xda,Kamada:2016euw}. In addition to the other problems, this {\it diversity} problem constitutes the small-scale crisis of CDM (see the recent reviews by \cite{Bullock:2017xww} and \cite{Tulin:2017ara}).

It has been suggested that these problems may be solved within the CDM framework by including baryonic processes such as star formation, supernova feedback and cooling into the $N$-body simulations~\citep{Navarro:1996bv,Governato:2009bg}, but whether this is sufficient in particular to solve the core-cusp problem is an ongoing debate~\citep{Read:2018fxs,Bose:2018oaj}. As a result, modifications of the assumed properties of DM have been considered. Self-interacting dark matter (SIDM) was first suggested by \cite{Spergel:1999mh} as another possibility to address the small-scale issues. The typical momentum-transfer cross section needed to solve these problems is of the order of magnitude $\sigma/m_{\text{DM}}\sim 1\ \text{cm}^2\text{g}^{-1}$.

At the same time bounds on the self-interaction cross section have been derived from a variety of different astrophysical systems, with the strongest ones arising from detailed studies of galaxy clusters.
Bounds resulting from the non-observation of offsets between DM and galaxy centroids as strong as $\sigma/m_{\text{DM}}\lesssim 0.47~\text{cm}^2\text{g}^{-1}$ have been claimed~\citep{Markevitch:2003at,Randall:2007ph,Kahlhoefer:2013dca,Harvey:2015hha}. Subsequently, these bounds have been shown to be weaker than initially estimated, $\sigma/m_{\text{DM}}\lesssim 2~\text{cm}^2\text{g}^{-1}$ ~\citep{Wittman:2017gxn} and are expected to be further diminished when taking into account the collisionality of galaxies~\citep{Kummer:2017bhr}. Stronger bounds are obtained when considering sub-halo evaporation~\citep{Markevitch:2003at,Randall:2007ph},  $\sigma/m_{\text{DM}}\lesssim 1~\text{cm}^2\text{g}^{-1}$. Based on core-sizes of galaxy clusters bounds as strong as $\sigma/m_{\text{DM}}\lesssim 0.1~\text{cm}^2\text{g}^{-1}$ have been claimed, but these bounds strongly depend on assumptions of the baryonic physics \citep{Kaplinghat:2015aga,2018ApJ...853..109E}. 

These results have all made the common assumption that a contact interaction between DM particles induces isotropic scattering. In this case the total cross section and the momentum-transfer cross section are very similar. Such models are relatively easy to implement in numerical studies, since self-interactions of this type have to be rare in order to satisfy observational constraints~\citep{Rocha:2012jg,Peter:2012jh}. However, in order to  address the small-scale problems and satisfy constraints from larger systems, it may be necessary to consider velocity-dependent self-interaction cross sections~\cite{Buckley:2009in,Loeb:2010gj,Kaplinghat:2015aga}, such that the scattering rate increases with decreasing velocity. Such a velocity dependence is in fact well-motivated from particle physics and arises for example in models with light mediators~\cite{Buckley:2009in,Feng:2009hw,Loeb:2010gj,Tulin:2012wi,Tulin:2013teo}. 

While the velocity-dependence itself can be included in numerical simulations in a straight-forward way~\citep{Vogelsberger:2012ku, Zavala:2012us}, it is no longer justified in these models to treat self-scattering as isotropic. In fact, for the case of a light mediator the differential scattering cross section is strongly peaked in the forward direction~\citep{Kahlhoefer:2013dca}. To implement such a model within the framework of rare self-interactions, it becomes necessary to introduce a cutoff angle and neglect the effect of small-angle scattering~\citep{Robertson:2016qef}. 

In this paper, we take a closer look at the effect of frequent small-angle scattering, which may give rise to a drag force and lead to the transport of energy~\citep{Yoshida:2000bx,Moore:2000fp,Gnedin:2000ea,Markevitch:2003at,Ackerman:mha,Feng:2009mn}. We develop a hybrid description for simulations of isolated haloes in the limit of frequent self-interactions. Using a hydrodynamical description we treat the DM particles as a collisionless fluid and implement the energy transport by solving a heat diffusion equation. We find that efficient heat conduction leads to a reduction of the central density of a DM halo, which can address the core-cusp problem. This approach is similar to the gravothermal fluid formalism that was developed to study globular clusters \citep{1980MNRAS.191..483L} and has subsequently also been applied to self-interacting DM~\citep{Balberg:2002ue,Ahn:2004xt,2011MNRAS.415.1125K,Pollack:2014rja,Essig:2018pzq,Choquette:2018lvq,Nishikawa:2019lsc}.

We present an implementation of such a hybrid description for the energy transfer due to frequent self-interactions in the $N$-body code \textsc{gadget-2} \citep{Springel:2005mi} using its formulation of smoothed particle hydrodynamics (SPH). As an application of this implementation, we study the evolution of cuspy DM haloes modeled with a Hernquist profile under the influence of frequent DM self-interactions. While we limit ourselves to systems with spherical symmetry in the present work, the framework can be generalised to non-symmetrical systems. Our implementation can be combined with existing implementations of rare self-interactions to perform simulations of SIDM that are valid for both large and small angle scatterings. This approach will enable robust predictions from simulations that will help to interpret observations more reliably.

This paper is organised as follows: In Sec.~\ref{Sec:Frequent}, we introduce frequent DM self-interactions. The implementation of energy transfer via frequent DM self-interactions is described in Sec.~\ref{sec:implementation}. The setup and initial conditions of our simulations are discussed in Sec.~\ref{sec:IC}. In Sec.~\ref{sec:testing}, we validate our simulations by  comparing to analytic predictions. The results of our simulations are presented in Sec.~\ref{sec:results} before we conclude in Sec.~\ref{Sec:Disc}.

\section{Frequent dark matter self-interactions}
\label{Sec:Frequent}
The strongest bound on the DM self-interaction cross section for velocity-independent scattering originates from the Bullet Cluster, more specifically from the requirement that the smaller DM halo (the subcluster) should not lose more that $20\texttt{--}30\%$ of its mass during core passage~\citep{Markevitch:2003at,Randall:2007ph}. Following~\cite{Kahlhoefer:2013dca}, this statement can be expressed in terms of the number of scattering events leading to the expulsion of a DM particle from the subcluster (so-called expulsive collisions). In this context, a collision between a DM particle from the main cluster with a particle from the subcluster is expulsive if the velocity of both outgoing particles exceeds the escape velocity $v_{\text{esc}}$ of the subcluster.

For a subcluster moving through a larger cluster with relative velocity $v_0$ the fraction of expulsive collisions is found to be $f=(1- v_{\text{esc}}^2/v_0^2)/(1+ v_{\text{esc}}^2/v_0^2)$ for the case of isotropic scattering. The fractional loss of DM particles can then be expressed as
\begin{align}
\frac{\Delta N}{N}=1-\exp\left(-\frac{\Sigma\sigma f}{m_{\text{DM}}} \right),
\end{align} 
where $\Sigma$ is the integrated background density and $\sigma/m_{\text{DM}}$ the self-interaction cross section over DM mass. 

In this approach, non-expulsive collisions are assumed not to affect the subcluster at all. This is a reasonable approximation if self-interactions are rare and $f$ is of order unity~-- a case that has been explored by a number of recent simulational studies~\citep{Robertson:2016xjh,Kim:2016ujt}. However, it is also possible to satisfy the bound from the Bullet Cluster for very large self-interaction cross sections provided that $f$ is sufficiently small. This is the case for example if the differential cross section is strongly peaked in the forward direction, as expected for interactions mediated by a light or massless exchange particle. In such a situation the majority of interactions do not lead to a significant transfer of energy and momentum, so that the total cross section (and hence the frequency of scattering processes) can be large without violating observational constraints.

For frequent self-interactions the majority of scattering processes do not lead to the immediate evaporation of DM particles, but they may affect the structure of the DM halo in different ways. To see this, consider a single DM particle, moving through a constant background density with velocity $\mathbf{v}_0$. After each scattering process with angle $\theta \ll 1$, the velocity in the direction parallel to $\mathbf{v}_0$ will be reduced by $\delta v_{\parallel} \approx v_0 \sin^2\theta$, whereas in the direction perpendicular to $\mathbf{v}_0$ the velocity will increase by $\delta \mathbf{v}_{\perp} \approx v_0 \sin \theta$. Over a large number of random scatters, $\delta \mathbf{v}_{\perp}$ will average to zero, but $\delta v_\parallel$ and $\delta v_{\perp}^2$ do not.

If we replace the single DM particle by an entire DM halo moving with velocity $v_0$, frequent self-interactions can therefore be thought of as causing a drag force:
\begin{align}
\frac{F_{\text{drag}}}{m_{\text{DM}}} \propto \frac{\sigma}{m_{\text{DM}}}\rho v_0^2 \; .
\end{align}
This drag force converts the directed motion of the DM halo into internal energy, i.e.\ random motion of the individual DM particles. Effectively, the DM halo is heated up, which can lead to the loss of particles via cumulative evaporation. This net effect of a large number of small-angle scatters is not captured when only considering the effect of rare scatterings with large momentum transfer.

In contrast to rare self-interactions, frequent self-interactions are notoriously difficult to implement in numerical simulations. The reason is that particles are typically expected to scatter multiple times during a single time-step, making an effective description of these interactions necessary. Indeed, since the mean free path $\lambda=(n\sigma)^{-1}$ is typically small ($\lambda\ll r$) for frequent self-interactions, we can approximately apply a fluid description. The idea is to model the energy transport with a heat diffusion equation for a pressureless fluid:
\begin{align}
\mathbf{j}=-\kappa\nabla T \; ,
\end{align}
where $\mathbf{j}$ denotes the heat flux and $\kappa$ the thermal conductivity. The temperature $T$ is assumed to be given by the one-dimensional velocity dispersion $\nu$ as $T=m_{\text{DM}}\nu^2/k_B$ (see e.g. \cite{Gnedin:2000ea}). Combined with energy conservation,
\begin{align}
\rho\frac{\text{d}u}{\text{d}t}=-\nabla \mathbf{j} \; ,
\end{align}
the diffusion equation takes the form
\begin{align}
\rho \frac{\text{d} u}{\text{d} t}=\mathbf{\nabla}(\kappa\mathbf{\nabla} T) \; ,
\label{eq:heat}
\end{align}
where $u$ is the internal energy per unit mass. As we will discuss below, the particle physics enters in the definition of the thermal conductivity $\kappa$, which depends in particular on the self-scattering cross section.

The idea to treat DM as a fluid has been used previously in the literature to study the structure~\citep{Gnedin:2000ea,Yoshida:2000bx,Moore:2000fp} and stability~\citep{Balberg:2002ue,Ahn:2004xt,2011MNRAS.415.1125K,Pollack:2014rja,Essig:2018pzq} of DM haloes for frequent self-interactions. The disadvantage of this approach compared to $N$-body simulations is that it cannot be directly extended to cosmological simulations and that one cannot easily include the effect of rare self-interactions with large momentum transfer, for which the effective description does not apply.

The present paper takes the first steps towards the inclusion of an effective description of frequent self-interactions in $N$-body simulations. We implement the resulting energy transfer in an existing code and validate our approach for a number of simple configurations. We discuss the scaling behaviour of the results and point out possible convergence issues. Our approach can then be combined with existing implementations of rare self-interactions, with the relative importance of the two effects depending on the underlying particle physics properties of the self-interaction cross section and the resolution of the simulation.

\section{Implementation of energy transfer}
\label{sec:implementation}

In this section we present our implementation of frequent DM self-interactions in the cosmological $N$-body code $\textsc{gadget-2}$ \citep{Springel:2005mi}. We adopt a hybrid description, in which DM is modeled as a collisionless fluid supplemented by an effective treatment of self-interactions. To calculate the energy transfer we use the SPH section of the code, which in fluid simulations is intended for computing the hydrodynamical forces. The central idea is to use a smoothing kernel $W(r, H)$ with kernel-support radius $H$ (i.e. the largest
$r$ for which $W(r)>0$) to calculate local averages of relevant quantities in a way that achieves high resolution and low numerical noise. For concreteness, the cubic spline kernel is given by \citep{1985A&A...149..135M}
\begin{equation}
W(r,H)=\frac{8\pi}{H^3}
\begin{cases}
1-6\left(\frac{r}{H}\right)^2+6\left(\frac{r}{H}\right)^3\,, & 0\leqslant\frac{r}{H}\leqslant\frac{1}{2}\,,\\
2\left(1-\frac{r}{H}\right)^3\,, & \frac{1}{2}<\frac{r}{H}\leqslant1 \,,\\
0\,, & \frac{r}{H}>1 \; .
\end{cases}
\end{equation}
A comparison of different kernel choices is provided in App.~\ref{app:kernel}.

For a given smoothing kernel the local density of DM particles $\rho$ is calculated as
\begin{equation}
\rho_i=\sum_{j}m_j W(|\mathbf{r}_{ij}|,H_i) \; .
\end{equation} 
We can then define mass- and kernel-averaged quantities as
\begin{align}
\bar{A_i} = \frac{1}{\rho_i}\sum_{j} A_i \, m_j \, W(|\mathbf{r}_{ij}|,H_i) \; .
\end{align}
Using this approach we can calculate the one-dimensional velocity dispersions $\nu$:
\begin{align}
\nu_{x,i}^2=\overline{\Delta v_{x,i}^2}-\overline{\Delta v_{x,i}}^2 \; ,
\end{align}
where $\Delta v_{x,i}=v_{x,i}-v_{x,j}$ is the relative velocity between the particle and its neighbouring particles in the kernel in the direction of interest. We calculate the velocity dispersion for the $x$-, $y$- and $z$-direction separately and obtain the total (3D) velocity dispersion $\nu_{3D}$ by summing the individual contributions. Assuming isotropy within the kernel, the radial velocity dispersion is simply given by $\nu_r^2=(1/3)\nu_{3D}^2$. 
The noise of $\nu_{r,i}^2$ is reduced by calculating the kernel average $\overline{\nu_{r,i}^2}$.

Frequent DM self-interactions cause a transfer of energy, which can be interpreted as a transport of the velocity dispersion $\nu$. We include this effect by solving the heat conduction equation given in Eq.~(\ref{eq:heat}). As shown by~\cite{Jubelgas:2004iw}, its discretised version for the implementation in SPH codes reads
\begin{align}
\frac{\text{d}u_i}{\text{d}t}=\sum_j\frac{m_j}{\rho_i\rho_j}\frac{(\kappa_j+\kappa_i)(T_j-T_i)}{|\mathbf{x}_{ij}|^2}\mathbf{x}_{ij}\nabla_i W_{ij} \; ,
\end{align}
where $u= \nu^2 / (\gamma -1) = k_B T/((\gamma-1)m_{\text{DM}})$ is the thermal energy per unit mass (the one-dimensional case was discussed before by~\cite{1985PASAu...6..207B}). With the replacement of $(\kappa_i+\kappa_j)/2$ by $\kappa_{ij}=(2\kappa_i\kappa_j)/(\kappa_i+\kappa_j)$ to ensure a continuous heat flux \citep{1999JCoPh.148..227C}, we obtain
\begin{align}
\frac{\text{d}u_i}{\text{d}t}=2(\gamma-1)\sum_j\frac{m_j}{\rho_i\rho_j}\frac{\kappa^{\prime}_{ij}(u_j-u_i)}{|\mathbf{x}_{ij}|^2}\mathbf{x}_{ij}\nabla_i W_{ij} \; ,
\end{align}
where $\kappa^{\prime}_{ij}=(m_{\text{DM}}/k_B)\kappa_{ij}$. 
The diffusion of the one-dimensional velocity dispersion is then given by
\begin{align}
\frac{\text{d}(\nu_i^2)}{\text{d}t}=2(\gamma-1)\sum_j\frac{m_j}{\rho_i\rho_j}\frac{\kappa^{\prime}_{ij}(\nu^2_j-\nu^2_i)}{|\mathbf{x}_{ij}|^2}\mathbf{x}_{ij}\nabla_i W_{ij} \; .
\label{equ:conduction}
\end{align}

At each time step this diffusion equation is solved for the radial velocity dispersion in order to obtain $\Delta \nu_{r,i}^2$, which corresponds to the kinetic energy transferred via frequent self-scatterings. The result, which can be positive or negative, then needs to be added to the squared velocity of particle $i$, ensuring energy conservation. Therefore, the velocity after energy transfer is given by $v_i^{\prime 2}=v_i^2+\Delta \nu_{r,i}^2$. Since thermal conduction is an effective description for energy transfer due to many scatterings for which the momentum transfers average out, the direction of the particle should not change, such that
\begin{align}
\frac{\mathbf{v}_i^{\prime}}{|\mathbf{v}_i^{\prime}|}=\frac{\mathbf{v}_i}{|\mathbf{v}_i|} \; .
\end{align}
Hence we obtain 
\begin{align}
\Delta \mathbf{v}_i=\left(\sqrt{\frac{v^{\prime 2}_i}{v^2_i}}-1\right)\mathbf{v}_i \; .
\end{align}
To effect such a change in velocity, we calculate the conduction acceleration $\mathbf{a}_i=\Delta \mathbf{v}_i/\Delta t$ for a given time-step $\Delta t$ and add it to the gravitational acceleration. As usual, we also add a limiter to ensure that $v^{\prime 2}_i$ can never be smaller than zero.

The remaining challenge is to obtain a prescription for calculating the thermal conductivity $\kappa$ in terms of the underlying physics, in particular the differential scattering cross section $\mathrm{d}\sigma / \mathrm{d}\Omega$. Usually $\kappa$ is calculated from the mean free path, which depends on the total cross section $\sigma$. However, a more appropriate quantity in the present context is the viscosity cross section
\begin{align}
\sigma_{\text{V}}=\int\sin^2\theta\frac{\text{d}\sigma}{\text{d}\Omega}\text{d}\Omega \; ,
\end{align}
as discussed e.g.~by~\cite{present1958kinetic,1964cpig.book.....M,1974cpig.book.....M,2008ApJ...678..950S} and also in the context of DM by~\cite{Tulin:2013teo,Cline:2013pca}. The viscosity cross section accounts for the fact that perpendicular scattering ($\theta=\pi/2$) is most efficient for heat transfer in a DM halo and for altering its density profile, while both forward and backward scattering have a reduced effect.\footnote{We note that for long-range interactions, the viscosity cross section is still logarithmically dependent on the small-angle cut-off, illustrating the fact that frequent interactions with small scattering angles dominate the energy transfer.}

In analogy to the momentum transfer mean free path $\lambda_{\text{M}}=(n\sigma_{\text{T}})^{-1}$ defined in \cite{1964cpig.book.....M}, we therefore define the heat transfer mean free path
\begin{align}
\lambda_{\text{H}}=\frac{1}{n \, \sigma_{\text{V}}}= \frac{m_\text{DM}}{\rho \, \sigma_{\text{V}}} \; , 
\end{align}
where $n$ is the number density. For a monatomic gas we have $\kappa \sim c_v \lambda_{\text{H}} n \nu$, where $c_v=3k_B/2$ is the specific heat per particle. We hence obtain for the conductivity
\begin{align}
\kappa\sim\frac{3 k_B \nu}{2 \sigma_{\text{V}}} & & \text{and} & &  \kappa^{\prime}\sim\frac{3 \nu}{2} \left(\frac{\sigma_{\text{V}}}{m_\text{DM}}\right)^{-1} \; . \label{eq:kappa_simple}
\end{align}
This treatment is supported by a detailed calculation in \cite{present1958kinetic} that shows that the viscosity $\eta$ and thus the thermal conductivity ($\kappa=5/2\eta c_v$ for monatomic gases) scale as $\eta\sim\sigma_{\text{V}}^{-1}$.

To obtain a more precise expression for $\kappa$ we can employ the gravothermal fluid formalism used to study thermal conduction from DM self-interactions \citep{Balberg:2002ue,Ahn:2004xt,2011MNRAS.415.1125K,Pollack:2014rja,Essig:2018pzq}. This formalism makes use of the flux equation
\begin{equation}
 \frac{L}{4\pi r^2} = - \kappa \frac{\partial T}{\partial r} = - \kappa' \frac{\partial \nu^2}{\partial r} \; ,
\end{equation}
where $L$ is the heat radiated through a sphere of radius $r$. Thus, we can obtain $\kappa'$ from the relation between $L$ and the radial derivative of the velocity dispersion, replacing the total cross section by the viscosity cross section~\citep{Boddy:2016bbu}.

We treat the short and long mean free path limits following \cite{1968MNRAS.138..495L} and \cite{Balberg:2002ue}.
Let us first consider the case that the mean free path $\lambda_\text{H}$ is much smaller than the gravitational scale height, i.e.\ the Jeans length $\lambda_\text{J}$ given by $\lambda_\text{J}^2=\nu^2/(4\pi G \rho)$. In this short mean free path regime the flux is given by 

\begin{align}
\frac{L}{4\pi r^2}=-\frac{3}{2}\frac{b}{\hat{a}}\rho\frac{\lambda_{\text{H}}^2}{t_r}\frac{\partial \nu^2}{\partial r} \; ,
\end{align}
where $t_r$ is the relaxation time. The constant $\hat{a}=\sqrt{16/\pi}\approx 2.257$ is the collision rate of DM particles (assuming a Maxwellian distribution) and $b=25\sqrt{\pi}/32\approx 1.385$ is derived from Chapman-Enskog theory~\citep{lifshitz1981physical}. The relaxation time can be written as $t_r=(\hat{a} \sigma_{\text{V}} \rho\nu /m_{\text{DM}})^{-1}$, leading to
\begin{align}
\frac{L}{4\pi r^2} = -\frac{3 b \nu}{2}\left(\frac{\sigma_{\text{V}}}{m_{\text{DM}}}\right)^{-1}\frac{\partial \nu^2}{\partial r} \; ,
\end{align}
which fixes the constant of proportionality in Eq.~(\ref{eq:kappa_simple}). 

If the heat transfer mean free path is larger than the Jeans length, particles can complete several orbits between two scatters in which significant energy is transferred. In this case the important length scale for conduction is the Jeans length $\lambda_\text{J}$ rather than the mean free path.  In this long mean free path regime, the flux is given by
\begin{align}
\frac{L}{4\pi r^2}&=-\frac{3}{2}C\rho\frac{\lambda_\text{J}^2}{t_r}\frac{\partial \nu^2}{\partial r}\\
&=-\frac{3}{2}\hat{a}C\left(\frac{\sigma_{\text{V}}}{m_{\text{DM}}}\right)\frac{\nu^3\rho}{4\pi G}\frac{\partial \nu^2}{\partial r} \; .
\label{equ:lmfp} 
\end{align}
The constant $C$ appearing in this equation must be determined by $N$-body simulations~\citep{2011MNRAS.415.1125K}. A value of $C\approx 0.75$ is suggested for an NFW profile, while for a Hernquist profile the suggested value is $C\approx 0.9$ ($C\approx 1$ according to \cite{Balberg:2002ue}). A very recent study finds $C = 0.6$ for an NFW profile \citep{Essig:2018pzq}. As we will consider a Hernquist profile, we adopt $C = 0.9$, emphasizing that uncertainties in $C$ enter linearly in the calculation of $\kappa^{\prime}$ in the long mean free path regime.

To interpolate between the short and long mean free path regimes, we take
\begin{align}
\kappa'=\left(\frac{1}{\kappa'_{\text{lmfp}}}+\frac{1}{\kappa'_{\text{smfp}}}\right)^{-1} \; ,
\end{align}
which yields
\begin{align}
\kappa^{\prime}_i=\frac{3}{2}\hat{a} b\nu_{r,i}\left(\frac{\sigma_{\text{V}}}{m_{\text{DM}}}\right)\left(\hat{a}\left(\frac{\sigma_{\text{V}}}{m_{\text{DM}}}\right)^2+\frac{b}{C}\frac{4\pi G}{\rho_i \nu_{r,i}^2}\right)^{-1} \; .
\end{align}

We demonstrate in Appendix \ref{app:one-dim} that our code is able to reproduce the analytic solution for a one-dimensional test problem.

\section{Simulating core formation in dark matter haloes}
\label{sec:IC}

As discussed above, DM self-interactions cause energy transfer within DM haloes from regions with high temperature (i.e.\ high velocity dispersion) to colder regions. This effect heats up central cusps (which are colder than their surroundings) and transforms them into isothermal cores. For the remainder of this paper we will study this core formation, i.e.\ the transition from a cuspy profile to a cored profile. We will determine how the core size depends on both the age of the system and the strength of the self-interactions and determine appropriate rescaling rules.

To initialise our simulations, we model individual DM haloes with cuspy Hernquist profiles~\citep{Hernquist:1990be}, for which the enclosed mass, density profile and one-dimensional velocity dispersion are given by
\begin{align}
M(r)=&M\frac{r^2}{(r+a)^2}\label{Equ:mass}\;,\\
\rho(r)=&\frac{M}{2\pi}\frac{a}{r}\frac{1}{(r+a)^3}\;,\\
\nu_r^2(r)
		     =&\frac{GM}{12a}\left(\frac{12r(r+a)^3}{a^4}\log\Bigg(\frac{r+a}{r}\right)\nonumber\\
		     &-\frac{r}{r+a}\left[25+52\frac{r}{a}+42\left(\frac{r}{a}\right)^2+12\left(\frac{r}{a}\right)^3\right]\Bigg)\;,
\end{align}
where $M$ is the total mass and $a$ is the scale radius. The $N$-body solver furthermore requires the introduction of a softening length, and we choose the Plummer equivalent softening length according to \cite{Power:2002sw}.
We use the tool \textsc{makegal} to generate the initial conditions for spherical haloes in dynamical equilibrium \citep{Springel:1998dk}. \textsc{makegal} first samples the mass profile using the analytic expression Eq.~(\ref{Equ:mass}). The velocities of the DM particles are drawn from random samples to sample the distribution function of the Hernquist profile.

We focus on a DM halo with $M=10^{10}~\text{M}_{\odot}$, corresponding to the size of a typical dwarf galaxy. The scale radius $a=6.4~\text{kpc}$ is chosen to satisfy the concentration-mass relation from \cite{Dutton:2014xda}.  
 We define the dynamical time as the orbital time at the scale radius as in \cite{Kochanek:2000pi,Robertson:2016qef}:
\begin{align}
t_{\text{dyn}}=4\pi \sqrt{\frac{a^3}{GM}} \; ,
\end{align}
implying $t_{\text{dyn}}=0.98~\text{Gyr}$ for the halo that we consider. We run simulations with $N=10^6$ and $N=10^7$ particles. For the smaller number of particles, numerical cores, which form due to gravitational force softening \citep{Robertson:2016xjh}, start to grow for $t> 5~t_{\text{dyn}}$. For the case of $N=10^7$ stability of the initial conditions is guaranteed for the entire timescale we simulate. Adding the conduction module to the SPH version of $\textsc{gadget-2}$ increases the runtime of the code by roughly 30 per cent.

Following \cite{2011MNRAS.415.1125K, 2012MNRAS.423.3740V} we also introduce a dimensionless cross section
\begin{align}
\hat{\sigma}&=\frac{\sigma_{\text{V}}}{m_{\text{DM}}} \frac{M}{a^2} \nonumber \\ 
 & = 0.05\left(\frac{\sigma_{\text{V}}/m_{\text{DM}}}{1\text{cm}^2\text{g}^{-1}}\right)\left(\frac{M}{10^{10}\text{M}_{\odot}}\right)\left(\frac{6.4~\text{kpc}}{a}\right)^2
\end{align}
and a dimensionless time
\begin{align}
t_0^{-1}&=t_{r,0}^{-1}=\hat{a}\frac{\sigma_{\text{V}}}{m_{\text{DM}}}\rho_0 v_0=\hat{a}\frac{\sigma_{\text{V}}}{m_{\text{DM}}}\sqrt{\frac{G M^3}{2\pi^2a^7}}\;,
\end{align}
where $\rho_0=M/(2\pi a^3)$ and $v_0 =a\sqrt{4\pi G \rho_0}$.

We find that for cross-sections $\hat{\sigma}\lesssim3$ the entire halo (down to $r \sim 0.01 a$, which are the smallest radii that can be resolved in the simulation) resides in the long mean free path regime for the timescales we simulate. This implies that the thermal conductivity $\kappa$ is directly proportional to the self-interaction cross section and hence we expect larger self-interactions to lead to larger core sizes and more rapid core formation. Finally, we note that in the long mean free path limit $\kappa$ depends on the combination $C \sigma_{\text{V}}$, where $C$ is an empirical parameter with a considerable uncertainty (see Eq.~(\ref{equ:lmfp}) and surrounding text). Results for different values of $C$ can be obtained from the results presented below by an appropriate rescaling of $\sigma_{\text{V}}$.

\begin{figure*}
\centering
\includegraphics[width = 1\columnwidth]{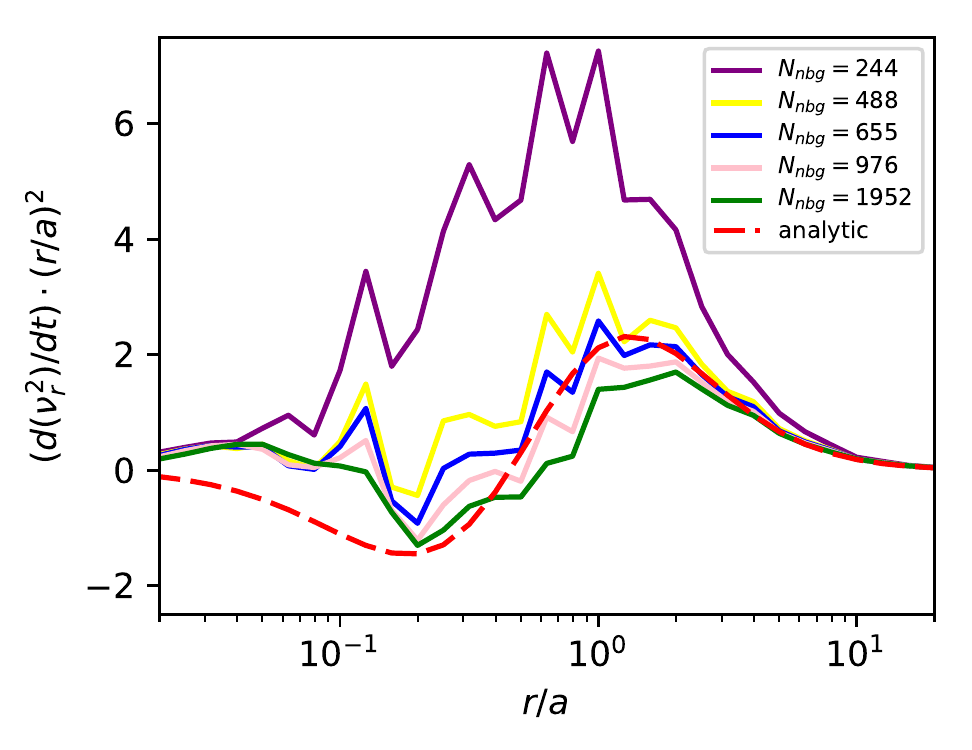}
\includegraphics[width = 1\columnwidth]{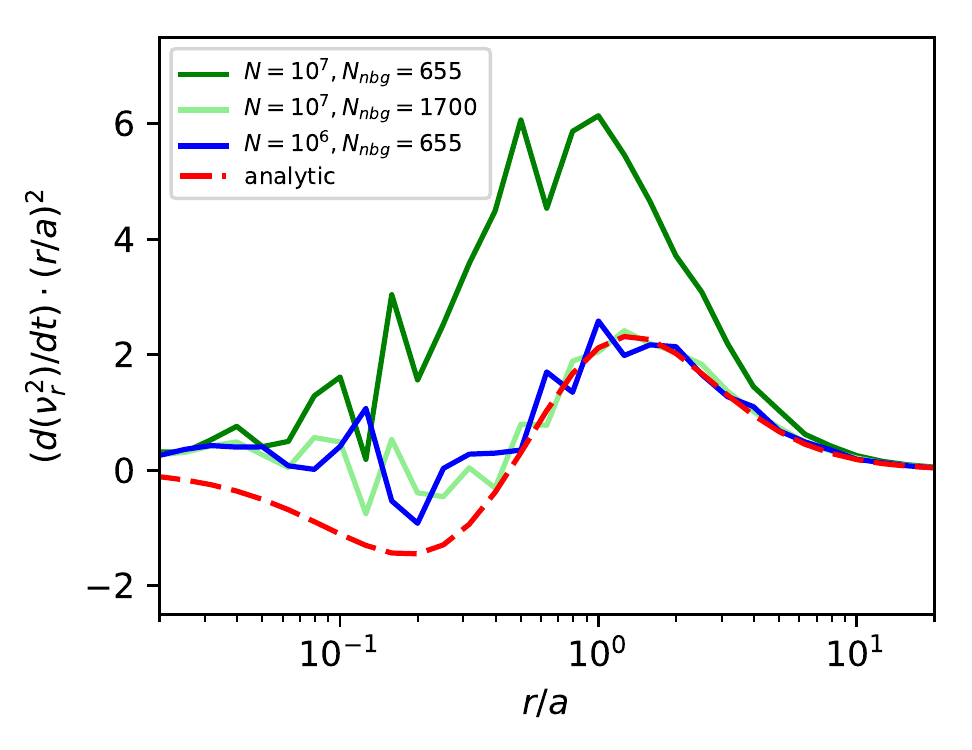}
\caption{Comparison of the median of the distribution of $\text{d}(\nu_r^2)/\text{d}t$ with the theoretical expectation for a Hernquist profile $(d(\nu_r^2)/dt)_{\text{th}}$ for $\sigma_{\text{V}}/m_{\text{DM}}=1~\text{cm}^2\text{g}^{-1}$, left panel: demonstration of increasing the number of neighbours in the cubic spline kernel with $N=10^6$. For $N_\text{nbg}=655$ we find the best agreement with the analytic expectation. Increasing $N_\text{nbg}$ further spoils this agreement. Right panel: Comparison for different total numbers $N$ of particles in the simulation. An increase of the number of particles $N$ does require an adjustment of $N_{\text{nbg}}$. The optimal value for $N=10^7$ is displayed ($N_\text{nbg}=1700$).}
\label{fig:convergence}
\end{figure*}

\section{Validation}
\label{sec:testing}

To validate the implementation presented above we compare the heat fluxes calculated by the code immediately after initialisation (i.e.\ before applying the first time-step) to the analytic expectation for a Hernquist profile. Specifically, we compare the respective rates $\text{d}(\nu_r^2)/\text{d}t$ obtained from the solution of the heat equation, see Eq.~(\ref{equ:conduction}). For the numerical result we calculate the median of all particles in a radial shell to reduce noise. The comparison with the analytic result is shown in Fig.~\ref{fig:convergence} for $\sigma_{\text{V}}/m_{\text{DM}}=1~\text{cm}^2\text{g}^{-1}$. In the left panel, we fix the total number of particles in the simulation to $N=10^6$ and consider different numbers of neighbours $N_\text{nbg}$ in the smoothing kernel, while in the right panel, both $N$ and $N_\text{nbg}$ are varied.\footnote{At first sight, the fact that $\mathrm{d}(\nu^2) / \mathrm{d}t$ is negative in the central region is counter-intuitive, given that we expect an increase in temperature in this region. The reason for this behaviour is that a self-gravitating system has negative heat capacity. In other words, reducing the kinetic energy of particles in the centre leads to an increase in total energy due to the virial theorem. This forces the particles onto larger orbits and drives the transformation of a cusp into a core~\citep{Pollack:2014rja}.}

We observe that for values of $N_\text{nbg}$ as typically chosen for SPH problems ($N_\text{nbg} \sim 10^2$) we substantially overestimate the conduction effect. The reason is that there are no additional hydrodynamical forces that reduce the noise in the velocity distribution. We thus need to increase the number of neighbours for the cubic spline kernel and thereby the kernel size to achieve a sufficient reduction in noise.\footnote{Note that we can allow a large number of neighbours even for the cubic spline kernel because we do not compute hydrodynamical forces and therefore are not limited by pairing instabilities~\citep{2012MNRAS.425.1068D}.} At the same time making $N_\text{nbg}$ too large partially averages out real physical effects. This reasoning implies the existence of an optimum range for $N_\text{nbg}$.

We estimate this optimum value using the least-squares method, i.e.\ by minimising
\begin{align}
    \chi^2 = \sum_j \frac{1}{\overline{\sigma}_j^2}\left[\overline{\text{d}(\nu_r^2)/\text{d}t}_j-\left(\text{d}(\nu_r^2)/\text{d}t\right)_{\text{th},j}\right]^2 \; ,
\end{align}
where $\overline{\text{d}(\nu_r^2)/\text{d}t}_j$ is the median of the numerical results in the $j$th radial shell and $(\text{d}(\nu_r^2)/\text{d}t)_{\text{th},j}$ is the theoretical prediction for a Hernquist profile. The uncertainty $\overline{\sigma}_j$ of the median is difficult to determine, since for neighbouring particles the values of $\text{d}(\nu_r^2)/\text{d}t$ are highly correlated. If $\sigma_j$ denotes the standard deviation of $\text{d}(\nu_r^2)/\text{d}t$ within each radial shell, we expect $\overline{\sigma}_j = \epsilon \, \sigma_j / \sqrt{N_{\text{sh},j}}$, where $N_\text{sh}$ is the number of particles in the radial shell and $\epsilon >1$ is some correction factor that accounts for correlations. Since we will only be interested in determining the best-fit value of $N_\text{nbg}$ and not in a goodness-of-fit estimate, we can drop this factor and simply determine the minimum of
\begin{align}
    \tilde{\chi}^2 = \sum_j \frac{N_{\text{sh},j}}{\sigma_j^2} \left[\overline{\text{d}(\nu_r^2)/\text{d}t}_j-\left(\text{d}(\nu_r^2)/\text{d}t\right)_{\text{th},j}\right]^2 \; .
\end{align}
%
For the fitting procedure we focus on a limited range around $r/a\sim 1$, which ensures on the one hand that there is a sufficiently large number of particles in each shell and on the other hand that the size of the smoothing kernel does not bias our results. For $N=10^6$ we achieve the best agreement with the analytic expectation in this range with $N_\text{nbg}=655$.
Increasing the total number of particles $N$ in the simulation induces more noise in the distribution of $\text{d}(\nu_r^2)/\text{d}t$. To compensate for this effect we must increase $N_\text{nbg}$ (see right panel of Fig.~\ref{fig:convergence}). We find $N_\text{nbg}$ to scale approximately proportional to $N^{0.4}$ and obtain an optimal vaule of $N_\text{nbg}=1700$ for $N=10^7$. We emphasize that numerical convergence in SPH is only possible in the simultaneous limit $N\rightarrow\infty$ and $N_\text{nbg}\rightarrow\infty$ \citep{Zhu:2014qka}. 

Moving to more extended kernels (such as Wendland C6 as proposed by \cite{2012MNRAS.425.1068D} or Wendland C8) does not improve the agreement compared to the cubic spline kernel, but leads to higher computational cost. Details of the kernel comparison are presented in App.~\ref{app:kernel}.

For very small radii, $r \ll a$ some discrepancies between analytical and numerical result remain even with the optimum choice for $N_\text{nbg}$. The reason is that the requirement of a large number of neighbours implies a large kernel-support radius $H$. Once $H$ becomes comparable to the radial distance of the particle to the centre of the halo, our implementation no longer produces reliable results, because the particles sampled by the kernel no longer provide unbiased estimates of the underlying distributions. In other words, it may no longer be possible to find a sufficient number of particles with smaller radius to reliably estimate the gradient in the velocity dispersion.

Fortunately, the limited resolution in the central region does not impede our ability to study core formation. The reason is that the core formation rate depends on the total amount of heat transported towards the centre. This heat transport can be accurately estimated from our implementation, which agrees well with the analytic expectations for $r \gtrsim a/2$. However, the numerical artefacts in the central region may lead to unphysical energy transport \emph{within} the core, which may affect core stability. We therefore cannot cover the full evolution of the halo in the gravothermal fluid model and do not expect to accurately capture the recollapse of cores at very late times \citep{1968MNRAS.138..495L,Balberg:2002ue,2011MNRAS.415.1125K,Pollack:2014rja,Kochanek:2000pi,Robertson:2016qef}.

\section{Results} 
\label{sec:results}

\begin{figure*}
\centering
\includegraphics[width = 1\columnwidth]{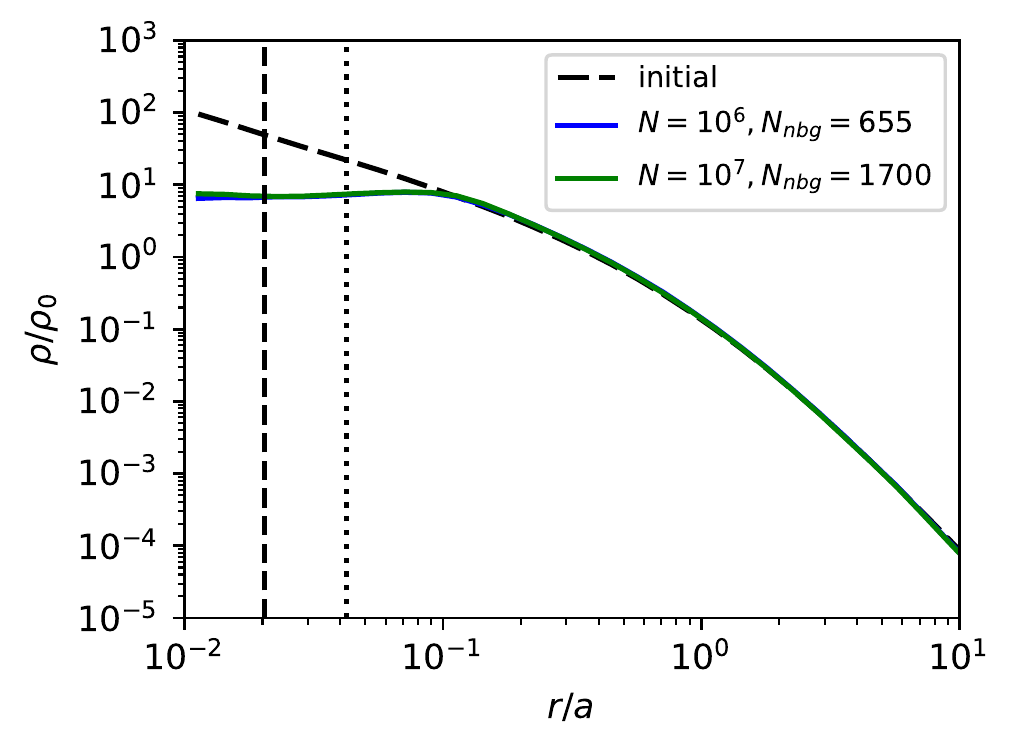}
\includegraphics[width = 1\columnwidth]{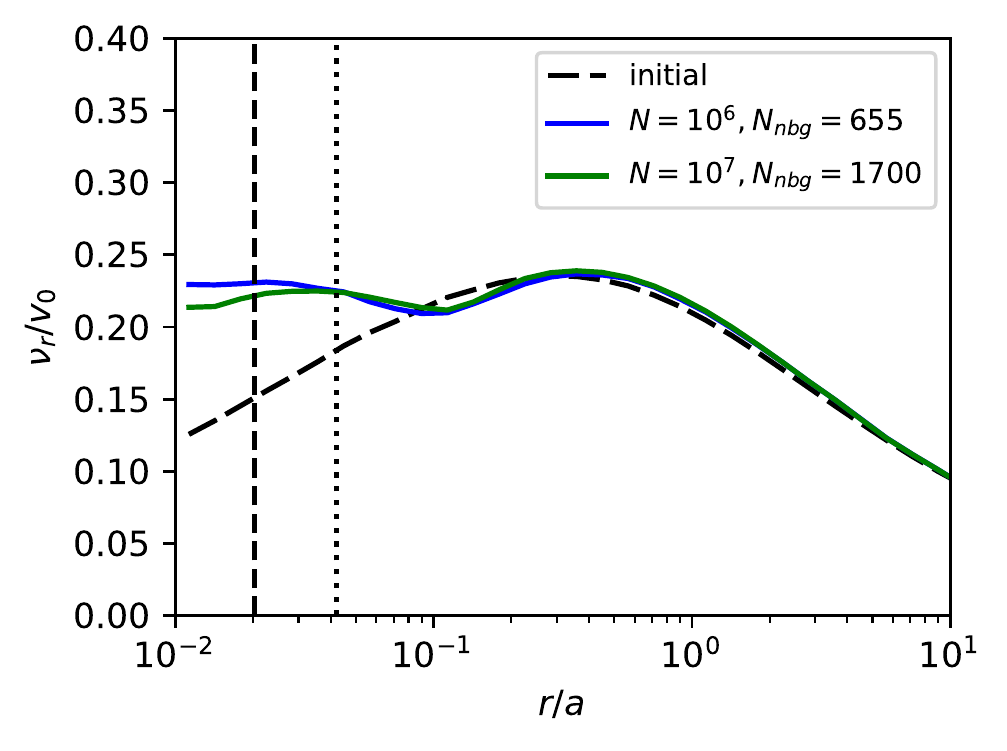}\\
\caption{Left panel: dimensionless density profile. Right panel: dimensionless velocity dispersion. We show both for $\sigma_{\text{V}}/m_{\text{DM}}=1~\text{cm}^2\text{g}^{-1}$ at $t\approx t_0$. The agreement of the density profiles for $N=10^6$ and $N=10^7$ shows that our method converges for the given setup. For numerical reasons the velocity dispersion is not exactly flat in the central region. The dashed vertical line indicates the softening length for $N=10^7$ and dotted for $N=10^6$.}
\label{fig:profiles}
\end{figure*}

\begin{figure}
\centering
\includegraphics[width = 1\columnwidth]{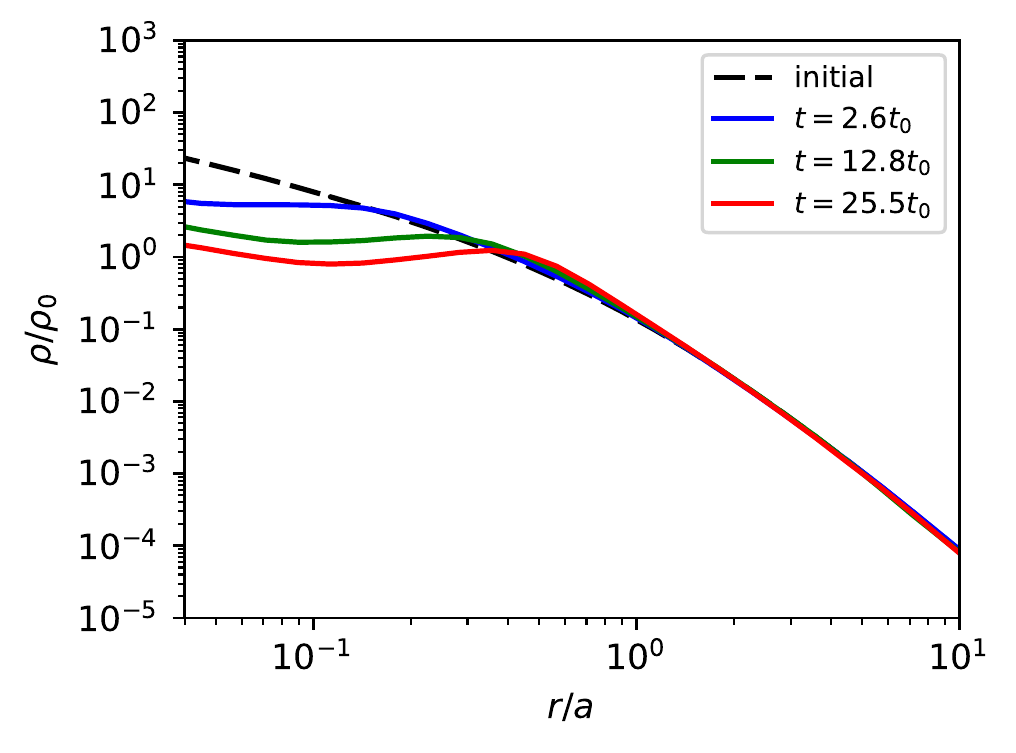}
\caption{Time evolution of the dimensionless density profile of the dwarf-size DM halo for $\hat{\sigma}=1$.}
\label{fig:time_evolution}
\end{figure}

Based on our findings in the previous section, we use the cubic spline kernel with $N_\text{nbg}=655$ for $N=10^6$ and $N_\text{nbg}=1700$ for $N=10^7$ to simulate the dwarf-size DM halo defined in Sec.~\ref{sec:IC}. Fig.~\ref{fig:profiles} shows the density profile and velocity dispersion of the halo after time $t_0$, defined as the relaxation time at the centre. In agreement with expectation we find heat conduction from frequent DM self-interactions to reduce the central density and to start core formation. At the same time the velocity dispersion flattens in the centre of the halo, leading to an isothermal profile. The fact that the velocity dispersion is not exactly flat in the centre stems from the numerical limitations discussed in Sec. \ref{sec:testing}. The results for $N=10^6$ and $N=10^7$ agree very well, indicating that our simulations converge for $\sigma_{\text{V}}/m_{\text{DM}}=1~\text{cm}^2\text{g}^{-1}$. On purpose, our fiducial choice of particles is not the maximal number that we can afford but  $N=10^6$ and $N_\text{nbg}=655$ because our technique is intended to be used in larger simulation boxes where only moderate particle numbers per halo can be afforded. A comparison to simulations with $N=10^7$ particles supports the presented results with a deviation of the final core size smaller than 2 per cent for $\hat{\sigma}=1$.


\begin{figure*}
\centering
\includegraphics[width = 1\columnwidth]{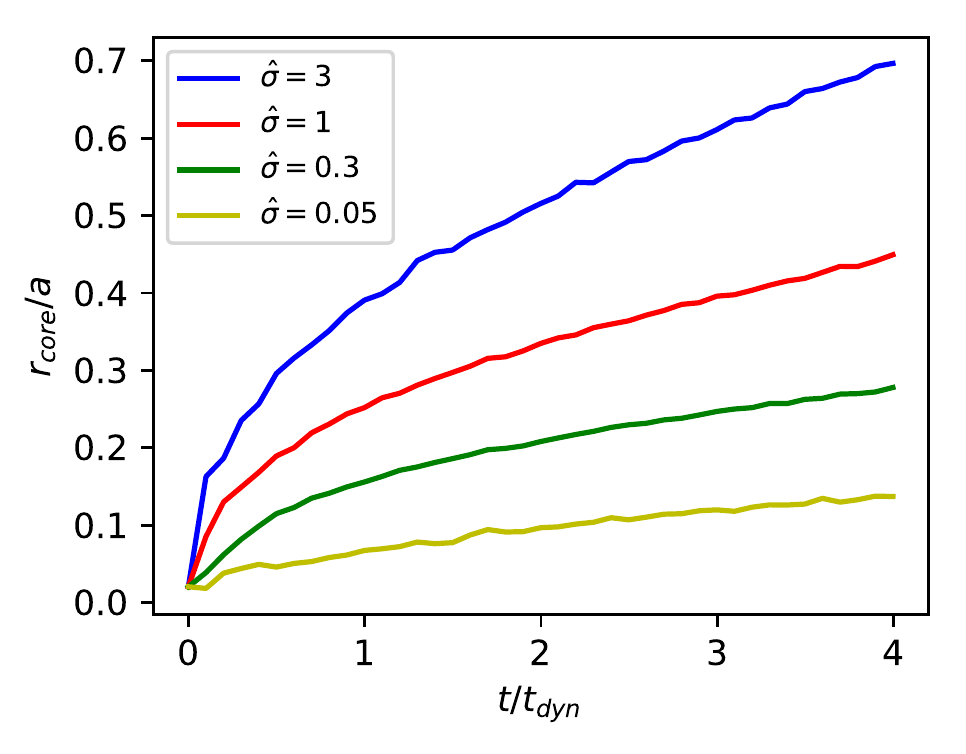}
\includegraphics[width = 1\columnwidth]{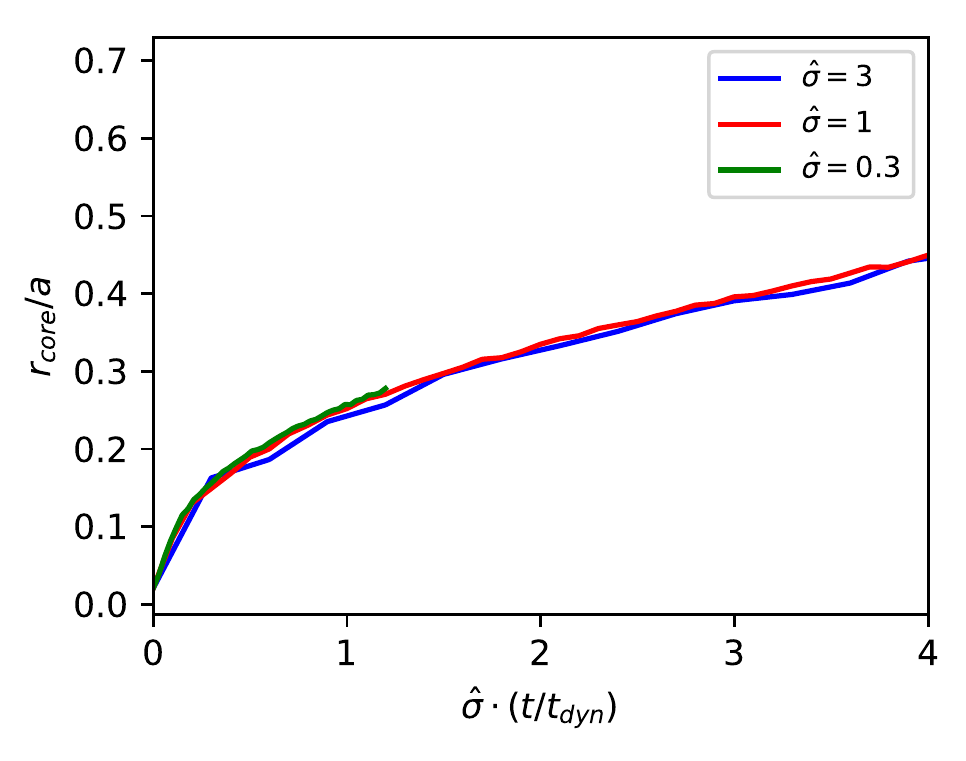}
\caption{Core growth for different values of the dimensionless cross section. In the right panel we rescaled the time with the dimensionless cross section. The agreement implies that the growth rate scales linearly with the cross section. The largest cross section is expected to be affected by the gravothermal catastrophe, which we do not cover in our simulation}
\label{fig:core-growth}
\end{figure*}

Fig.~\ref{fig:time_evolution} illustrates the time evolution of the DM halo in dimensionless units for $\hat{\sigma}=1$ after $t=2.6~t_0$, $t=12.8~t_0$ and $t=25.5~t_0$. 
We extract the core size of the DM halo by fitting the density profile to a cored Hernquist profile
\begin{align}
\rho(r)=&\frac{M}{2\pi}\frac{a}{(r^\beta+r_{\text{core}}^{\beta})^{1/\beta}}\frac{1}{(r+a)^3} \; .
\end{align}
We keep the parameter $\beta=4$ fixed and allow the other parameters to vary during the fitting processes \citep{Robertson:2016qef}. 
The core growth for different dimensionless cross sections $\hat{\sigma}$ is shown in the left panel of Fig.~\ref{fig:core-growth}. 
The growth of the core is very rapid in the beginning and decreases with time.
For comparison, \cite{Pollack:2014rja} found a maximal core size of a similar magnitude for an NFW profile as we find at $t\approx25~t_0$ with the same cross section\footnote{Note the difference in the definition of the dimensionless cross section in \cite{Pollack:2014rja} $\hat{\sigma}^{\prime}=\hat{\sigma}/(4\pi)$ such that $\hat{\sigma}^{\prime}=0.08$ is equivalent to $\hat{\sigma}=1$.}.

In the right panel of Fig.~\ref{fig:core-growth} we show the core size as a function of the product of dimensionless time and cross section. The fact that all curves lie on top of each other implies that the growth rate depends linearly on the cross section. This scaling originates from the fact that the entire halo is in the long mean free path limit (see discussion in the end of Secs.~\ref{sec:implementation} and~\ref{sec:IC}).

We find that the core size at $t=4~t_\text{dyn}$ is $r_{\text{core}}/a\approx 0.45$ for $\hat{\sigma}=1$. This is in agreement with the results of \cite{Kochanek:2000pi} as they find a maximal core size of $r_{\text{core}}/a\approx 0.4$ for the same cross section. This core size already reached at $t\approx t_\text{dyn}$, in our case the evolution is much slower. Note that the maximal core size found by \cite{Robertson:2016qef} is much smaller ($r_{\text{core}}/a\approx 0.15$) compared to our results and the results of \cite{Kochanek:2000pi}. 

Since we are not able to resolve the very centre of the halo, we are not able to simulate the full evolution of the gravothermal fluid model. Especially for large cross sections a recollapse of the core is expected for the timescale we simulate. For example, for $\hat{\sigma}=3$ we have $t_0\approx0.05~\text{Gyr}$, which means we cover $t=4\,t_{\text{dyn}}\approx80 \, t_0$ in our simulation, while \cite{2012MNRAS.423.3740V} found that the recollapse starts after $t\approx25\,t_0$. The prediction of the core size for the large cross section $\hat{\sigma}=3$ is therefore not reliable, since we expect the core to recollapse. Note that $\hat{\sigma}=3$ corresponds to a very large cross section of $\sigma_{\text{V}}/m_{\text{DM}}\approx59~\text{cm}^2\text{g}^{-1}$ which is not compatible with current bounds. For cross sections which are compatible with the bounds and for timescales smaller than the age of the Universe recollapse is not relevant for isolated DM haloes.

\section{Discussion}
\label{Sec:Disc}
We have presented the  first $N$-body simulations using SPH to capture the effect of DM self-interactions for frequent scattering.
Using an implementation of thermal conduction in the SPH section of \textsc{gadget}-2 we carried out simulations of an (isolated) dwarf-sized DM halo. To map the particle physics parameters to the thermodynamical quantities (i.e. the thermal conductivity) we use the well-established idea of the gravothermal fluid formalism.
The resulting energy transfer from the outer part of the halo towards the centre has a significant impact on the evolution of the system. The central density of an initially cuspy halo is reduced and an isothermal core is created. We therefore recover the well-known mechanism for solving the core-cusp problem and the too-big-to-fail problem with SIDM.

Due to the negative heat capacity of self-gravitating systems, any self-interacting DM halo will experience a gravothermal catastrophe~\citep{1968MNRAS.138..495L} at very late times or for very large cross sections~\citep{Balberg:2002ue,Kochanek:2000pi}. The fact that the resulting DM haloes would be in obvious conflict with observations can be used to obtain relevant constraints on the self-interaction cross section, in particular when including effects from dissipation~\citep{Essig:2018pzq}. While it would be very interesting to investigate these effects in our simulations, numerical limitations in the central region of the DM halo imply that it is currently not possible to simulate its recollapse. It remains to be seen whether this problem can be overcome with higher-resolution simulations or different smoothing kernels.

Our formalism provides a proof of principle for the inclusion of frequent DM self-interactions in $N$-body simulations and should be seen as a first step towards a comprehensive treatment of all effects of SIDM. For example, our simulations do not presently include the effective drag force created by frequent self-interactions, i.e. we simulate the SIDM effects on the halo via heat conduction only. The reason for neglecting this drag force is that it is expected to be small in the system with spherical symmetry, as the velocity distribution of an isolated halo and therefore in each kernel is isotropic. In order to study anisotropic systems such as merging galaxy clusters, an implementation of an effective drag force will be needed.

Furthermore, the set-up of our simulations does not fully capture the relevant physical process during structure formation. Considering initially cuspy profiles implicitly assumes that self-interactions are not relevant for structure formation, which may not be a good approximation for large self-interaction cross sections. In order to extract reliable predictions and bounds from our formalism it will be essential to perform full cosmological simulations. The fact that such simulations involve many different length scales complicates the optimum choice of the smoothing kernel and may necessitate adaptive kernel sizes.

Finally, we have focussed on simulations of DM only. Understanding the interplay of baryonic physics and SIDM is an important next step, in particular in the context of the core-cusp problem. For simulations of rare DM self-interactions there has been a lot of progress on this subject recently. Testable predictions of SIDM have been developed and a preference for small self-interaction cross-sections ($\sigma/m_\text{DM}\approx0.1~\text{cm}^2\text{g}^{-1}$) from core sizes in galaxy clusters has been found \citep{2018ApJ...853..109E,Robertson:2018anx,2018MNRAS.479..359S,Despali:2018zpw} although baryonic feedback introduces a sizeable scatter~\citep{Robertson:2017mgj}.

It is important to emphasize that our approach for simulating frequent self-interactions should not be seen as an alternative to existing state-of-the-art simulations of SIDM. Indeed, our ultimate goal is to combine the implementation presented here with the well-established formalism for rare scattering in order to simulate self-scattering cross-sections with arbitrary angular dependence. The idea would be to explicitly simulate large-angle scatters while using our approach to capture the effect of small-angle scatters. This hybrid solution will then make it possible to perform fully consistent $N$-body simulations of observationally favoured SIDM models with scattering cross sections depending on velocity and scattering angle. Doing so will enable us to draw robust conclusions on the nature of DM self-interactions.


\label{sec:discussion}

\section*{Acknowledgements}
We thank Manoj Kaplinghat and Denis Wittor for valuable discussions. This work is funded by the Deutsche Forschungsgemeinschaft (DFG) through the SFB grant 676, the Emmy Noether Grant No.\ KA 4662/1-1,  the ERC Starting Grant `NewAve' (638528) and under Germany's Excellence Strategy -- EXC 2121 ,,Quantum Universe`` -- 390833306.



\bibliographystyle{mnras}
\bibliography{lit}


\appendix
\section{Kernel comparison}
\begin{figure}
\centering
\includegraphics[width = 1\columnwidth]{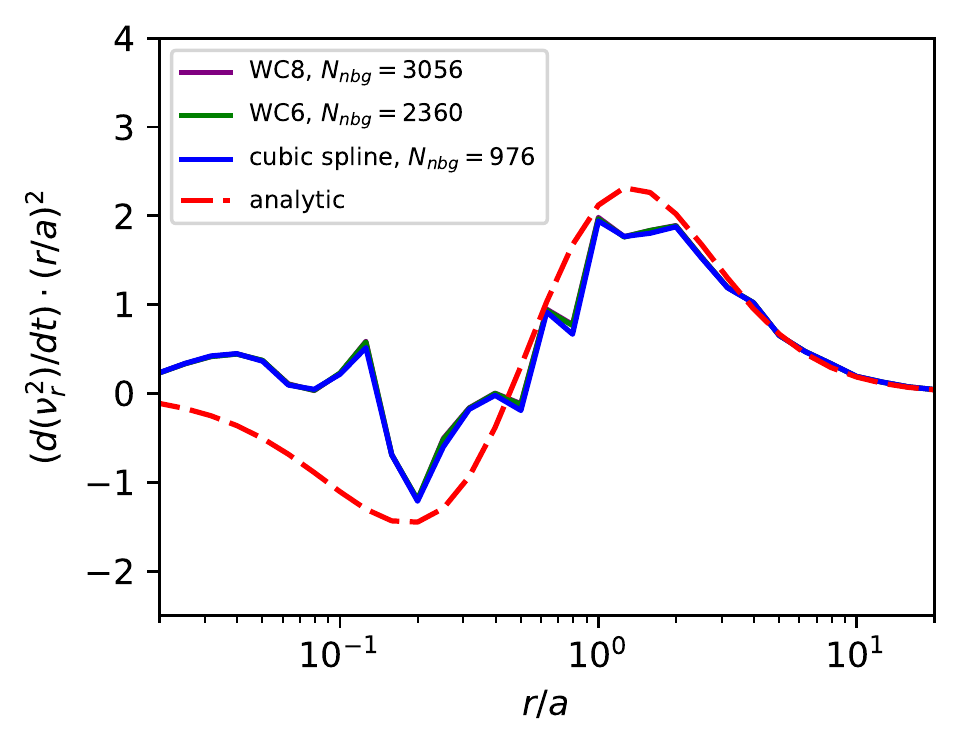}
\caption{Comparison of the median of the distribution of $\text{d}(\nu_r^2)/\text{d}t$ with the theoretical expectation for a Hernquist profile $(d(\nu_r^2)/dt)_{\text{th}}$ for $\sigma_{\text{V}}/m_{\text{DM}}=1~\text{cm}^2\text{g}^{-1}$. Here we compare different choices of the smoothing kernel with $N_\text{nbg}$ chosen such that the smoothing scale $h$ agrees for all kernels with $N=10^6$. We demonstrate that more extended kernels do not lead to any further improvement.}
\label{fig:convergence_app}
\end{figure}
We have experimented with different choices for the SPH kernel in our simulation. In particular, we have investigated the effect of more extended smoothing kernels, such as the Wendland C6 kernel (WC6) \citep{2012MNRAS.425.1068D} defined by
\begin{align}
&W(r,H)=\frac{1365}{64 \pi H^3}\left(1-\left(\frac{r}{H}\right)\right)^8_+\cdot\\
&\left(1+8\left(\frac{r}{H}\right)+25\left(\frac{r}{H}\right)^2+32\left(\frac{r}{H}\right)^3\right),
\end{align}
where $(\cdot)_+=\operatorname{max}\{0,\cdot·\}$.
More extended kernels obviously have the property of a larger effective kernel size (kernel-support radius $H$ i.e. the largest value of the radius,
$r$, for which $W(r)>0$). This increases the number of neighbour particles, while keeping the same smoothing scale $h$ \citep{2012JCoPh.231..759P}. The smoothing scale is essentially defined by the kernel standard deviation $h=2\sigma$. A typical choice for the WC6 kernel is $N_\text{nbg}=295$. Finally, we have also experimented with a WC8 kernel \citep{Wendland} defined as
\begin{align}
&W(r,H)=\frac{357}{64 \pi H^3}\left(1-\left(\frac{r}{H}\right)\right)^{10}_+\cdot\\
&\left(5 + 50 \left(\frac{r}{H}\right) + 210 \left(\frac{r}{H}\right)^2 + 450 \left(\frac{r}{H}\right)^3 + 429 \left(\frac{r}{H}\right)^4\right).
\end{align}
Keeping the smoothing scale, $h$, of the WC6 kernel with $N_\text{nbg}=295$ fixed, the WC8 kernel allows for $N_\text{nbg}=382$. We find that this choice leads to no significant improvement compared to the cubic spline kernel for the agreement of $\text{d}(\nu_r^2)/\text{d}t$ with the theoretical expectation $(d(\nu_r^2)/dt)_{\text{th}}$ as shown in Fig.~\ref{fig:convergence_app}. However, the computational cost increases substantially for the same smoothing length, $h$. Consequently, we need to increase the smoothing scale at the cost of losing resolution. This is achieved by increasing the number of neighbors for a fixed kernel. A useful quantity for the efficiency of the kernel is the number of neighbours inside the smoothing scale $N_{h}=(h/H)^3N_\text{nbg}$. $N_{h}$ is larger for the cubic spline kernel compared to the more extended kernels for a fixed number of particles. To keep the computational cost low we choose the cubic spline kernel for the simulations presented in Sec.~\ref{sec:results}. To summarise, we learned that the smoothing scale of the kernel needs to be increased by raising the number of neigbours $N_\text{nbg}$ for a given kernel in order to reduce the noise effectively.
\label{app:kernel}


\section{One-dimensional test problem}
\begin{figure*}
\centering
{\includegraphics[width = 0.675\columnwidth]{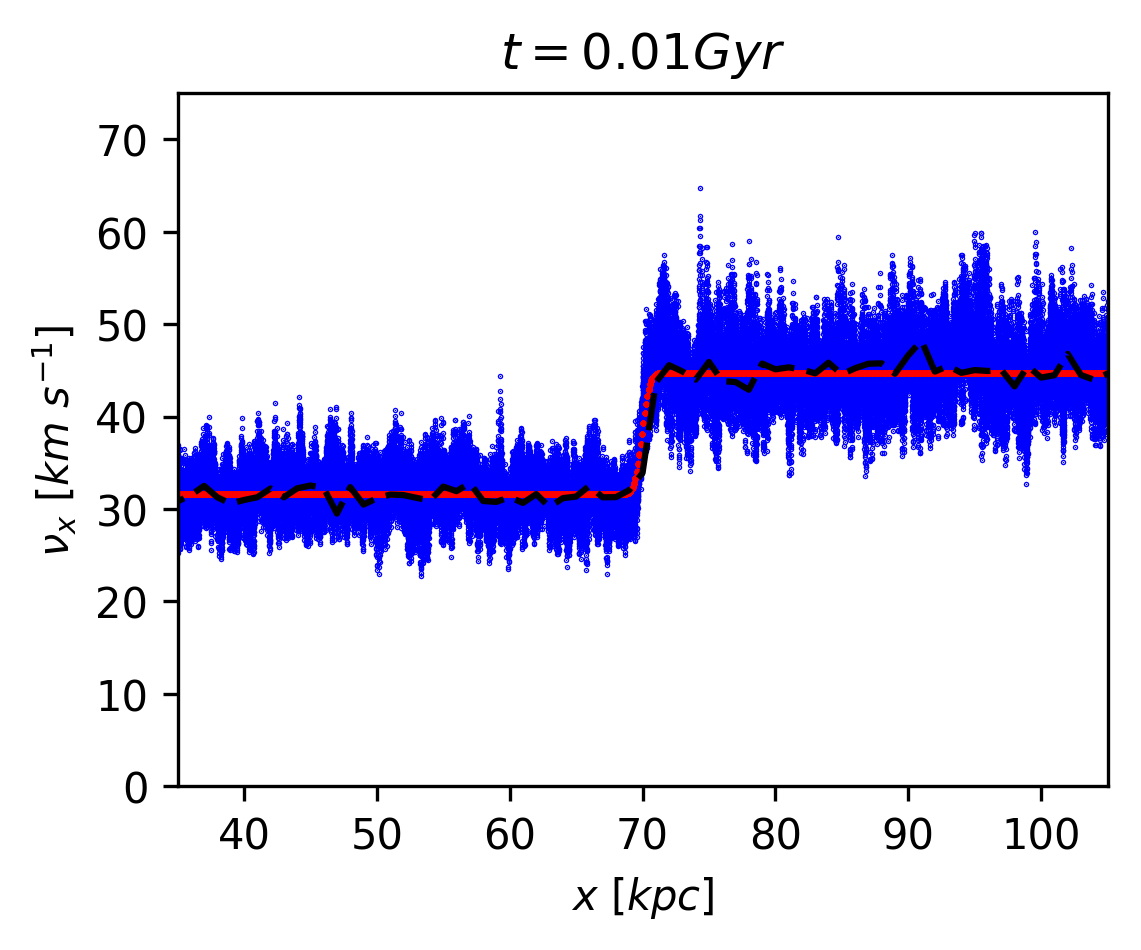}} 
{\includegraphics[width = 0.675\columnwidth]{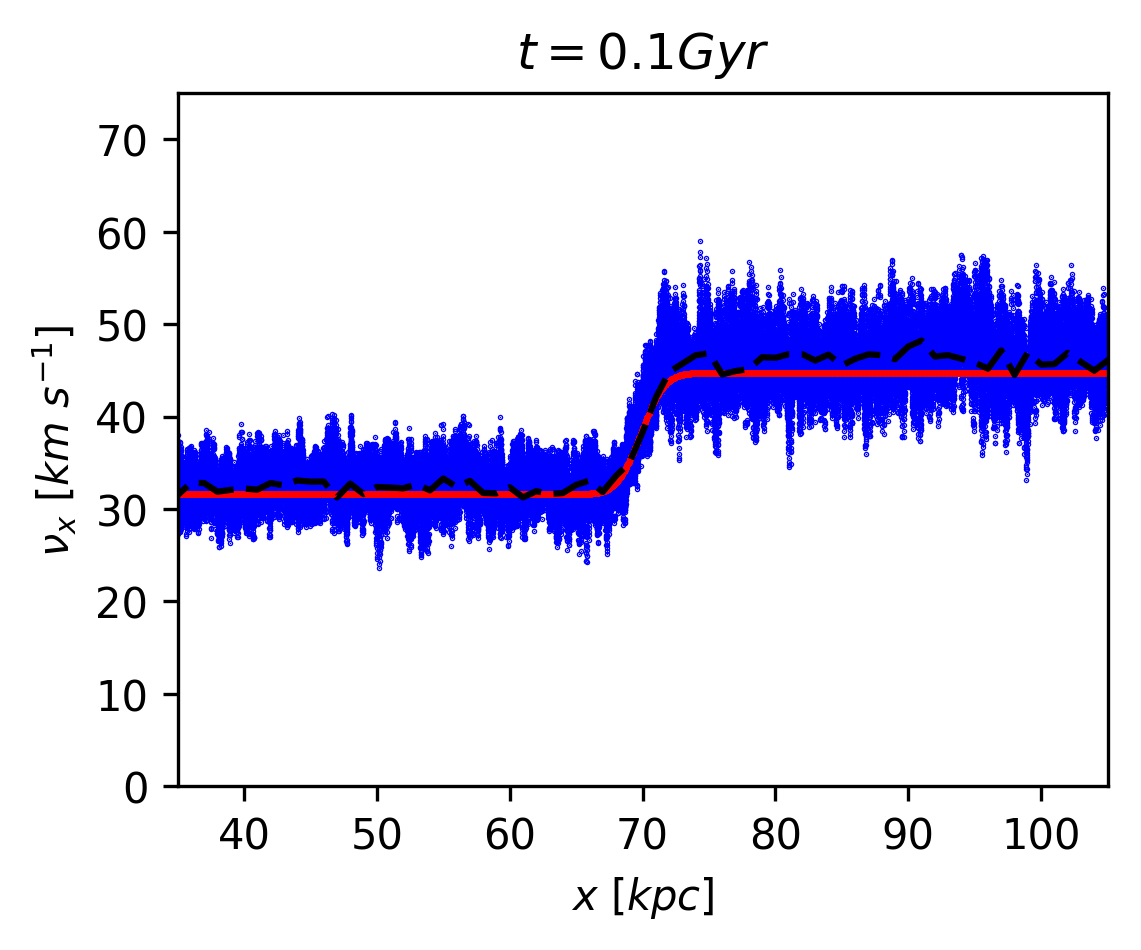}}
{\includegraphics[width = 0.675\columnwidth]{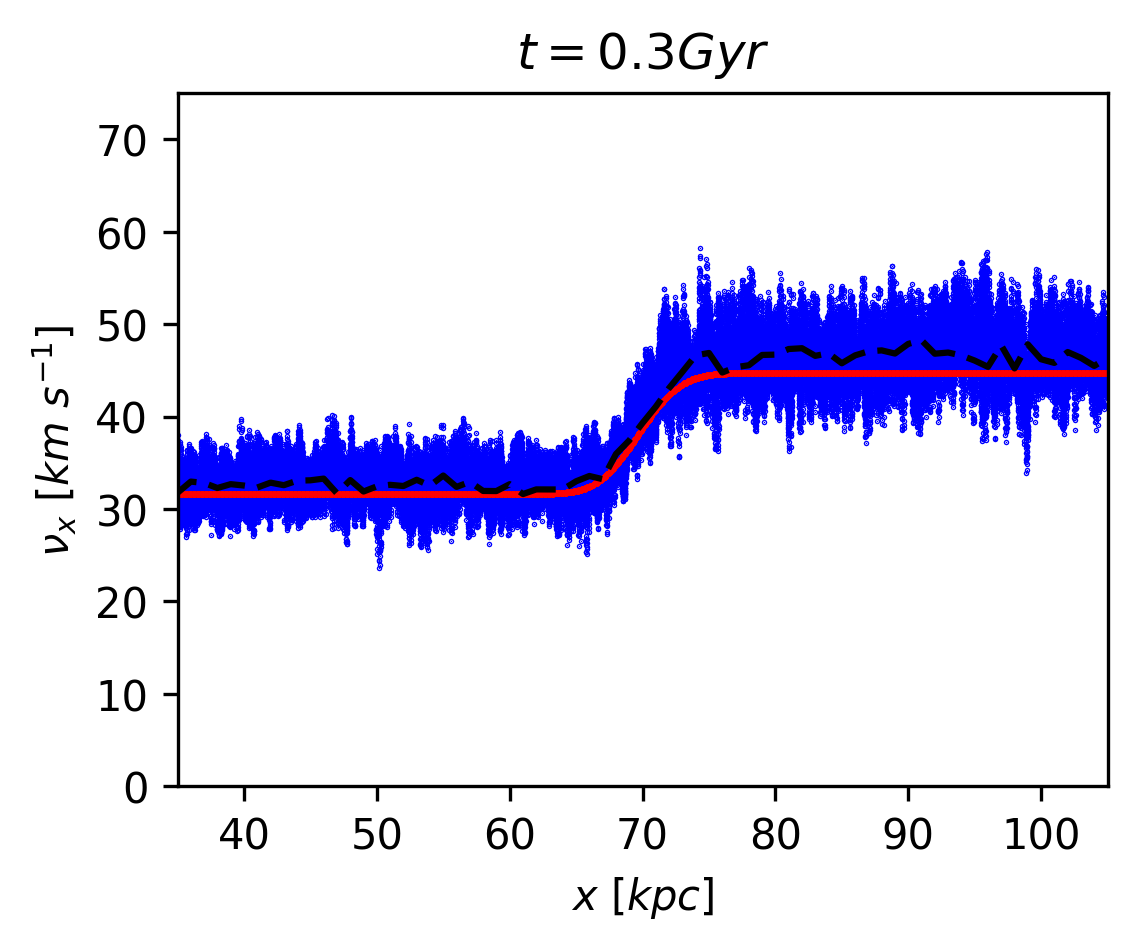}} \\ 
{\includegraphics[width = 0.675\columnwidth]{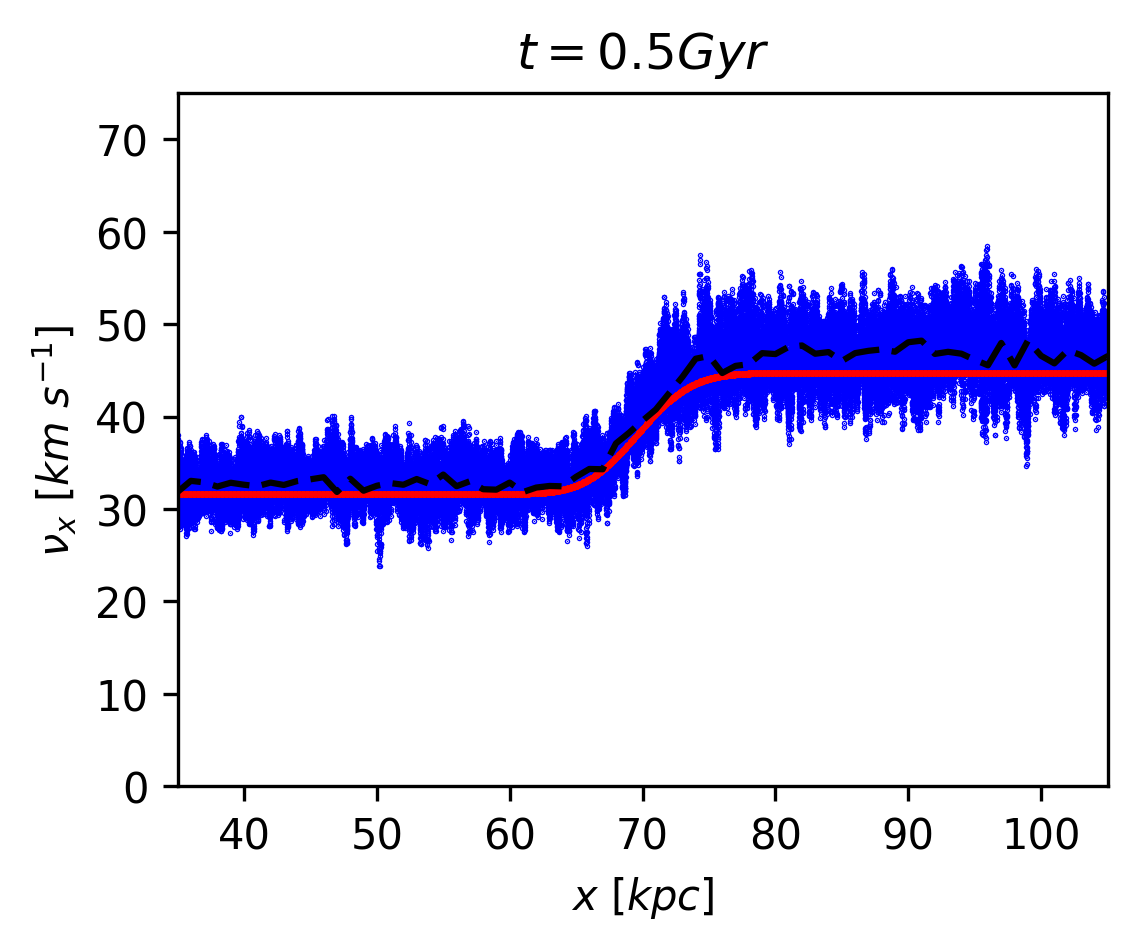}} 
{\includegraphics[width = 0.675\columnwidth]{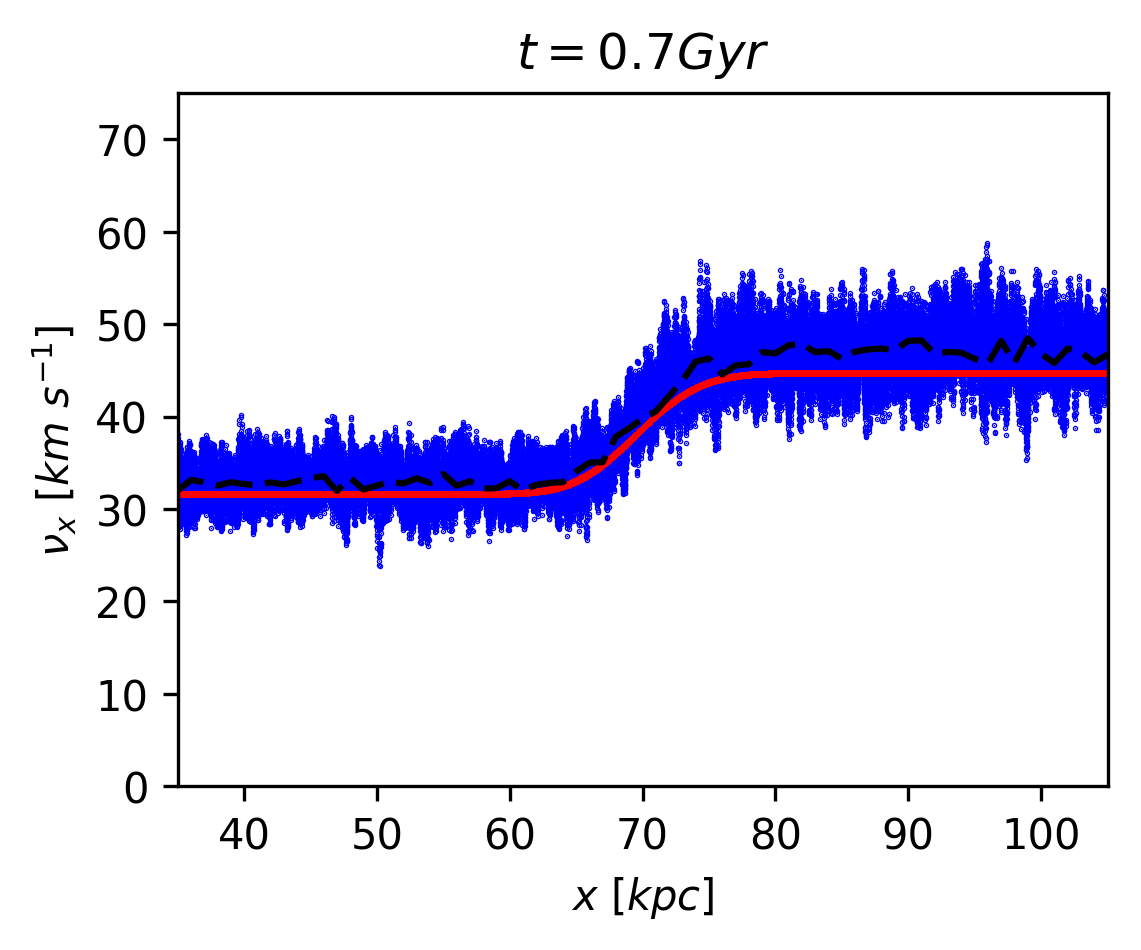}} 
{\includegraphics[width = 0.675\columnwidth]{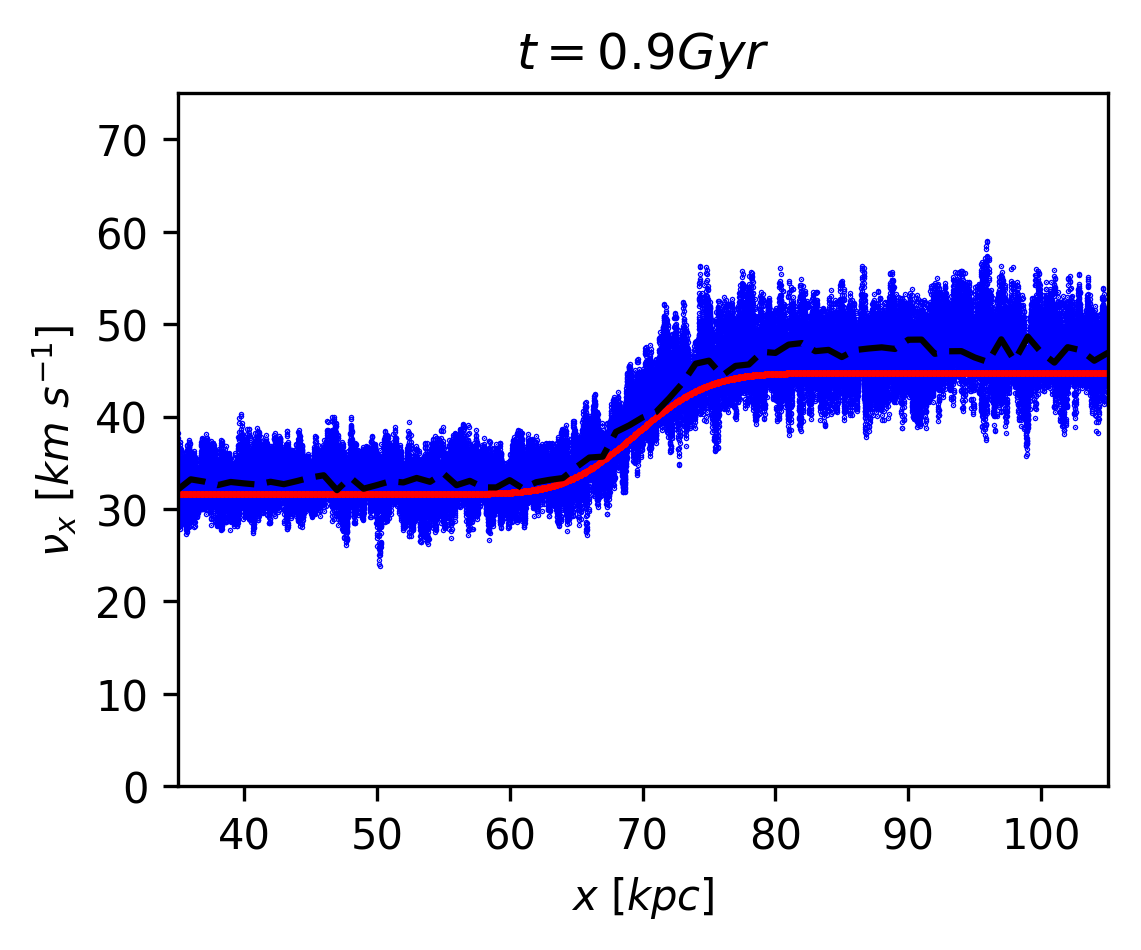}}\\
\caption{Time evolution of the velocity dispersion for the one-dimensional test problem. We show the analytic solution in red and the mean of the velocity dispersion distribution in black.}
\label{fig:1d-test}
\end{figure*}
For the purpose of testing our implementation of heat conduction, we set up a one-dimensional test without gravity and compared the results to the analytical solution. To this end, we initialised particles on a grid with random velocities but different velocity dispersions on either side of the centre of the computational domain. We calculate the velocity dispersions from the initial velocities. The velocities are updated owing to the conduction and, hence, the velocity dispersions change accordingly. However, in this particular test we did not allow the particles to move but instead kept them fixed on the grid. This way, we can study only the transport of heat and compare the results to the analytic solution. Initially, we set the conductivity $\kappa^{\prime}$ to a certain value and keep it constant throughout the whole test $\kappa^{\prime}=\alpha=10$ since $\alpha=\kappa^{\prime}/\rho$ and we chose $\rho=1$ in internal units. Converted to physical units, this implies $\rho=6.77\cdot10^{-22}~\text{g}/\text{cm}^3$, $\kappa^{\prime}=2.04\cdot10^{6}~\text{g}/(\text{cm}\cdot\text{s})$ and $\alpha=3.02\cdot10^{27}~\text{cm}^2/\text{s}$. For this case, the analytic solution is known \citep{Jubelgas:2004iw}
\begin{align}
\nu(x,t)=\nu_0+\frac{\Delta\nu}{2}\operatorname{erf}\left(\frac{x-x_0}{\sqrt{4\alpha t}}\right), 
\end{align}
where $x_0$ is the position of the initial difference of size $\Delta \nu$ of the velocity dispersion, and $\nu_0$ the mean velocity dispersion. We use the WC6 kernel with $N_\text{nbg}=295$ and find that the resulting thermal conduction of the simulation agrees well with the analytic solution around the centre for constant $\kappa^{\prime}$. The results are shown in Fig.~\ref{fig:1d-test}.
\label{app:one-dim}
\newpage


\bsp	
\label{lastpage}
\end{document}